\pdfoutput=1
\documentclass[sigconf]{acmart}
\AtBeginDocument{%
  \providecommand\BibTeX{{%
    \normalfont B\kern-0.5em{\scshape i\kern-0.25em b}\kern-0.8em\TeX}}}

\copyrightyear{2022} 
\acmYear{2022} 
\setcopyright{acmcopyright}\acmConference[JCDL '22]{The ACM/IEEE Joint Conference on Digital Libraries in 2022}{June 20--24, 2022}{Cologne, Germany}
\acmBooktitle{The ACM/IEEE Joint Conference on Digital Libraries in 2022 (JCDL '22), June 20--24, 2022, Cologne, Germany}
\acmPrice{15.00}
\acmDOI{10.1145/3529372.3530928}
\acmISBN{978-1-4503-9345-4/22/06}

\newtheorem{definition}{Definition}
\newtheorem{example}{Example}

\begin{document}

\title[Benefits of Narrative Information Access in Digital Libraries]{What a Publication Tells You - \\Benefits of Narrative Information Access in Digital Libraries}


\author{Hermann Kroll}
\email{kroll@ifis.cs.tu-bs.de}
\orcid{0000-0001-9887-9276}
\affiliation{%
  \institution{Institute for Information Systems, TU Braunschweig}
  \streetaddress{Mühlenpfordtstr. 23}
  \city{Braunschweig}
  \state{Lower Saxony}
  \country{Germany}
  \postcode{38106}
} 

\author{Florian Plötzky}
\email{ploetzky@ifis.cs.tu-bs.de}
\orcid{0000-0002-4112-3192}
\affiliation{%
  \institution{Institute for Information Systems, TU Braunschweig}
  \streetaddress{Mühlenpfordtstr. 23}
  \city{Braunschweig}
  \state{Lower Saxony}
  \country{Germany}
  \postcode{38106}
}  

\author{Jan Pirklbauer}
\email{j.pirklbauer@tu-bs.de}
\affiliation{%
  \institution{Institute for Information Systems, TU Braunschweig}
  \streetaddress{Mühlenpfordtstr. 23}
  \city{Braunschweig}
  \state{Lower Saxony}
  \country{Germany}
  \postcode{38106}
}

\author{Wolf-Tilo Balke}
\email{balke@ifis.cs.tu-bs.de}
\orcid{0000-0002-5443-1215}
\affiliation{%
  \institution{Institute for Information Systems, TU Braunschweig}
  \streetaddress{Mühlenpfordtstr. 23}
  \city{Braunschweig}
  \state{Lower Saxony}
  \country{Germany}
  \postcode{38106}
}   

\renewcommand{\shortauthors}{Kroll et al.}

\begin{abstract}
Knowledge bases allow effective access paths in digital libraries.
Here users can specify their information need as graph patterns for precise searches and structured overviews (by allowing variables in queries).
But especially when considering textual sources that contain narrative information, i.e., short stories of interest, harvesting statements from them to construct knowledge bases may be a serious threat to the statements' validity.
A piece of information, originally stated in a coherent line of arguments, could be used in a knowledge base query processing without considering its vital context conditions.
And this can lead to invalid results.
That is why we argue to move towards narrative information access by considering contexts in the query processing step.
In this way digital libraries can allow users to query for narrative information and supply them with valid answers.
In this paper we define narrative information access, demonstrate its benefits for Covid 19 related questions, and argue on the generalizability for other domains such as political sciences.
\end{abstract}



\maketitle

\section{Introduction}
\begin{figure*}
    \centering
    \includegraphics[trim=0.0cm 4.0cm 0.0cm 0.0cm, width=0.85\textwidth]{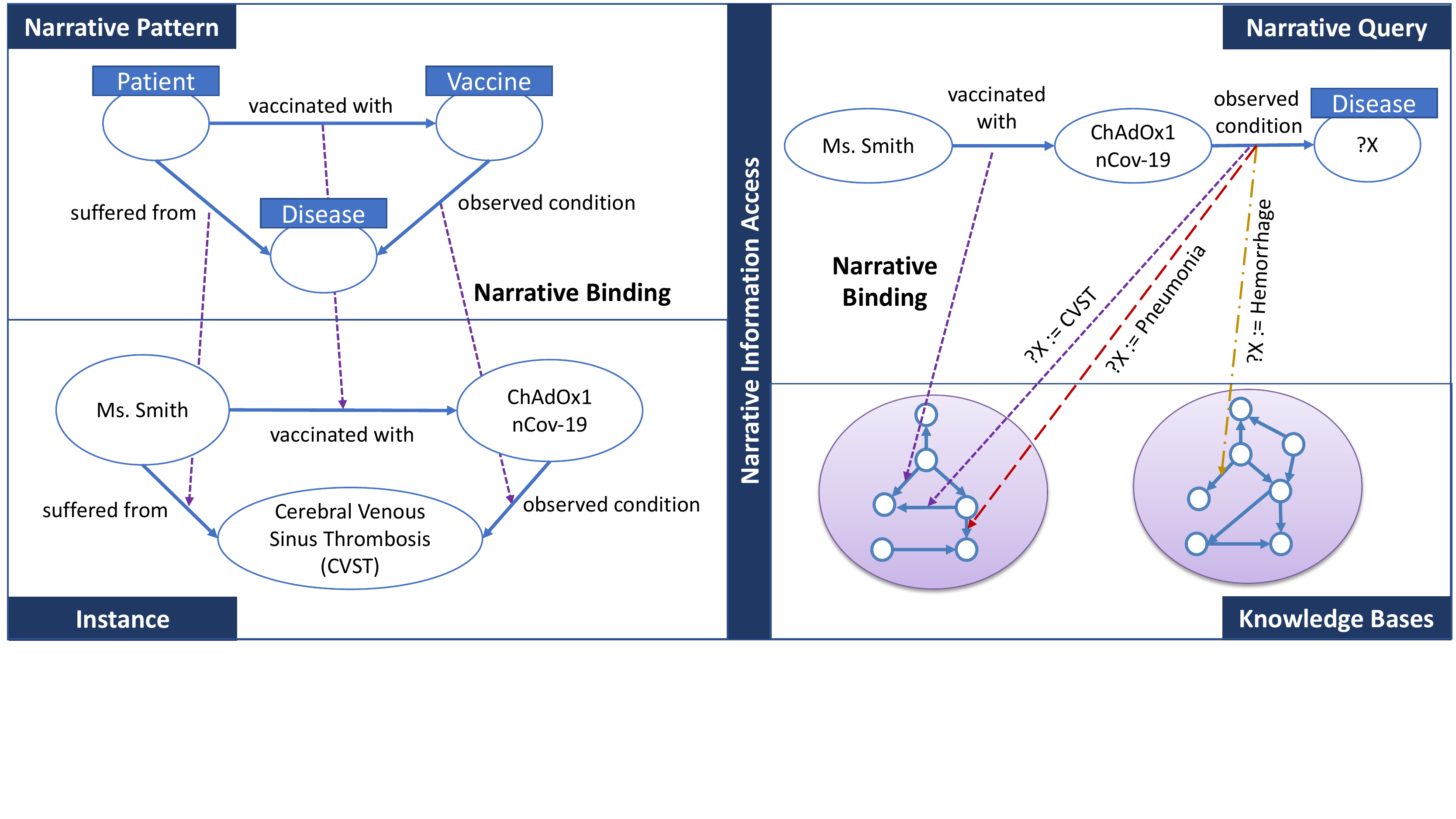}
    \caption{Systematic overview: A narrative pattern (upper left  corner) describes a template how different entity types interact with each other. An instance then substitutes the entity types by concrete entities (lower left corner). These substitutions are called narrative bindings. On the right, the narrative query processing is depicted: Narrative bindings are found for each statement of a narrative query. Bindings that share the same context are depicted in the same colour and shape.}
    \label{fig:narrativeinformationaccess}
\end{figure*}

From the beginnings of human language, knowledge was shared and passed on following a narrative oral tradition, i.e., they exchange stories and have structured debates and conversations~\cite{laszlo2008science}. 
With the advent of written language, these oral presentations were made persistent by writing up stories, comments and discussions in articles and books.
The central way to encode all this knowledge is to tell a story: a narrator relates what was observed and how more complex conclusions were derived from basic claims.
We thus understand this process as  \textit{composing narratives}, i.e., action patterns bound to real-world entities or concepts to form rich lines of arguments~\cite{freeman2011argument}.

Today digital libraries play a key role in making knowledge publicly available in large-scale repositories. 
The necessary curation builds on a long-standing library sciences tradition and results in a variety of novel digital technologies to manage and access knowledge repositories, including the FAIR principles~\cite{wilkinson2016fairprincles}:
On the one hand, extensive collections need to be effectively maintained and efficiently archived. 
Here additional metadata enrichment is already used in each source to prepare the data for later access (Findability \& Accessibility).
On the other hand, digital libraries face an increasing amount of data collected from distributed sources. 
This is done either by providing unifying interfaces to individual collections of linked open data or by using information integration techniques over extractions from different sources (Interoperability \& Reuse).

The traditional solution is to provide a simple keyword-based access path to the underlying data.
Then users have to retrieve this data and \textit{determine} what is actually \textit{told} by the data.
What happens here is that users try to \textit{understand} the data to reuse the information of interest for their purposes.
We understand this \textit{exploratory process of understanding} as \textit{gradually composing narratives}, in the sense of extracting and generalizing patterns that are 'told' by the data.
Take, for instance, the COVID-19 pandemic: 
Patient records might describe suffered conditions after they have been vaccinated by a SARS-CoV-2 vaccine. 
Biomedical experts can then read through these records and extract typical story patterns, e.g., patients may experience headaches and pain, or even worse, may suffer from dangerous cerebral sinus venous thrombosis.
Although this manual workflow is common, the rapid speed of the COVID-19 pandemics has shown that, given the amount of data available, it is hard to stay up-to-date.
Even when restricting information sources only to well-curated ones, researchers would have to cover nearly 200k peer-reviewed articles about COVID-19 published in the US National Library of Medicine over the last two years\footnote{\url{https://www.ncbi.nlm.nih.gov/research/coronavirus/}}. 

Such rapid developments ask for novel and more efficient access methods.
For example, a comprehensive database of all possible conditions observed in COVID-19 vaccinations might be helpful for improved diagnostics.
Yet, when building such a knowledge base by harvesting statements about COVID-19 from textual sources, the answer quality may not be sufficient in practice.
This is because the \textit{observed conditions} are torn from the original course of vaccination as exhibited by some concrete patient.
For example, some conditions might only be observed in elderly patients and thus, might not apply to children, or some complications might only be possible when a certain pre-existing condition is present in a patient.
This means that although each condition was correctly extracted, the reusing of the resulting statements in a knowledge base may not be valid because the information's \textit{contexts} do not match.
When humans read through publications and retrieve arguments, they usually consider all essential context conditions such as the treated group or relevant pre-existing conditions.
Moreover, in addition to contexts, humans also consider the connection between statements within a line of argument, e.g., do the assumptions within the arguments leading to a conclusion actually make sense together?

We argue that digital libraries need to move towards narrative information access, i.e., to offer query capabilities in the form of narrative patterns while considering vital contexts. 
Therefore we first define narrative information access. 
We then argue on contexts and how digital libraries can retain them. 
In addition, we perform two case studies on top of our narrative retrieval system, published last year~\cite{kroll2021narrativequerygraphs}. 
We investigate COVID-19-related research questions in cooperation with domain experts. 
We also asked an expert from the political sciences domain to study the system and describe how the political sciences domain could benefit from such a retrieval system.
Finally, we discuss the generalizability, benefits, and challenges of narrative information access for digital libraries.

\section{Narrative Information Access}

In the following section we define the concept of narrative information access and discuss its key components.
To ease understanding, we start with a running example from the biomedical field as a narrative pattern: Covid 19 vaccinations and their possible side effects. 
Consider the following short narrative:

\begin{example}
\label{exmp:pattern}
Some patients that were vaccinated by ChAdOx1 nCov-19 Vaccine (also known as Astra Zeneca) suffered Cerebral Venous Sinus Thrombosis (CVST).
Hence Intracranial Sinus Thrombosis is an observed disease condition for the ChAdOx1 nCov-19 vaccine.
\end{example}

Three types of entities participate in this example: a vaccine, patients, and a disease condition.
In addition, three possible relations between the entity types are expressed: patients \textit{are vaccinated} with the vaccine, patients \textit{suffer from} a disease condition, and the disease condition \textit{is observed} for the vaccine.
Thus narrative patterns are described by typing their participants and naming their relations (see  Fig.~\ref{fig:narrativeinformationaccess}).
The following ideas are based on an eased version of a narrative model that we introduced in~\cite{kroll2020er}.

Based on the encoding of knowledge in the Resource Description Framework (RDF)~\cite{manola2004rdf}, we define narrative patterns by:

\begin{definition}[Narrative Pattern]
A narrative pattern is a connected, node- and edge-labeled directed graph, where each edge (labeled with a predicate name) represents a statement in the form of a  (subject, predicate, object)-triple. Each node either represents a subject reflecting some entity type or an object reflecting either an entity type or literal values from a certain domain.
\end{definition} 

Any knowledge base in RDF format can then be seen as a graph containing a collection of \textit{instances of} narrative patterns as subgraphs, i.e., all nodes have been instantiated (either by URIs in the case of entities or by concrete literal values). 
We can now translate our previous example narrative using a narrative pattern as a kind of skeleton for the narrative. 
A possible instance is depicted in Fig.~\ref{fig:narrativeinformationaccess} (please note that for simplification, we replaced long URI prefixes with short entity names).

In brief, we have a graph representation of a concrete narrative structured by some narrative pattern. 
Hence narrative patterns can be understood as (sub-)graphs isomorphisms on RDF knowledge bases. 
We then define narrative queries using such patterns:

\begin{definition}[Narrative Query]
A narrative query is a narrative pattern where each node is either instantiated by a concrete entity or literal value or replaced by a variable (labeled by a variable name).
\end{definition} 

By design our proposed querying method has very similar semantics to querying RDF knowledge bases with SPARQL: 
If a narrative query does not contain a variable, then the answer is whether there exists an instance in the knowledge base that is isomorphic to the query's narrative pattern and features all the query's exact entities/literal values in the right places (cf. ASK queries in SPARQL).
If a narrative query contains one or more variables, then these variables must be substituted by concrete entities from the knowledge base during query processing. 
Of course, all matches to the query must be valid with regard to variable substitutions, i.e., the substituted pattern and the respective entities/values must be contained in the knowledge base. 
We understand such a matching process as \textit{binding a query}~\cite{kroll2021narrativebindings}, i.e., we take some edge of the query's narrative pattern and bind it against a knowledge base edge and bind concrete entities and literal values to the respective entity types or literal domains in the pattern.

Returning to our example, we may query which \textit{disease conditions} the \textit{ChAdOx1 nCov-19} vaccinated patient \textit{Smith} could possibly suffer from. 
The respective narrative query is depicted in Fig.~\ref{fig:narrativeinformationaccess}.
The first step to answer this query is to compute narrative bindings against the underlying knowledge base(s).
We may find a binding $b_1$ confirming that \textit{Ms. Smith} has been vaccinated with \textit{ChAdOx1 nCov-19}.
In addition, we must \textit{substitute} the variable \textit{?X (of type disease)}.
Here we may find three bindings with suitable substitutions: $b_2$ (CVST), $b_3$ (Pneumonia), and $b_4$ (Hemorrhage). 
In common graph querying we would now join the intermediate results to list all conditions that \textit{Ms. Smith} could possibly expect: CVST, pneumonia and hemorrhage.

Now, assume for the time being that \textit{pneumonia} have only been observed in elderly people, whereas \textit{Ms. Smith} is still young. 
Then \textit{pneumonia} as a possible side effect of the vaccination might no longer apply to \textit{Ms. Smith}, although the respective binding observing \textit{pneumonia} as a possible side effect of a \textit{ChAdOx1 nCov-19} vaccination is perfectly correct.
The problem here is that $b_3$ would not be valid \textit{in general}, because the observed conditions do not apply to all patients, but only to elderly patients.
Although the bindings are correctly retrieved, not all of them might actually fit into the context of \textit{Ms. Smith}.

Here information was torn apart regarding a sensitive context such as the target group information.
One might argue that extracting RDF-style knowledge from individual patient records could even in the best case be problematic and should not be done in this way.
While we agree that all patients are somewhat unique cases, this kind of extraction is common practice in real life applications, e.g., the \textit{causes} relation in SemMedDB~\cite{kilicoglu2012semmeddb}, \textit{medical causes} in Wikidata~\cite{vrandevcic2014wikidata}\footnote{\url{https://www.wikidata.org/wiki/Property:P828}}, and \textit{causes} in DBpedia~\cite{auer2007dbpedia}\footnote{\url{https://dbpedia.org/property/causes}}.

The effect is that even if knowledge bases did only contain correct statements, fusing them to answer a query may still produce incorrect results.
Indeed, it is a good scientific practice to arrange statements as complex lines of arguments, i.e., authors are sure to mention all essential contexts, settings, assumptions made, necessary conditions, hypotheses, experimental designs, etc.
It is essential to fuse only those arguments fitting into the same context provided in the form of constraints by other arguments or the query terms.
We call bindings \textit{context-compatible} if they can safely be fused to form valid knowledge.
Based on the idea of context-compatibility, we are now ready to propose a novel query processing method that considers contexts as constraints upon the query process to bypass the previous issues.

\begin{definition}[Narrative Query Processing]
Given a narrative query and a set of knowledge bases, the query processing has to a) bind each individual query statement against underlying data of the knowledge base(s) and b) check the context-compatibility of the computed bindings. The result of the query process is thus a set of valid bindings, individually binding all query statements and being context-compatible. 
\end{definition}

Thus narrative query processing ensures that contexts are considered while matching graph patterns. 
All bindings must in this way share a compatible context. 
And with this narrative query processing method we can now define narrative information access:

\begin{definition}[Narrative Information Access]
Narrative Information Access allows users to formulate their information need as a narrative query. A narrative retrieval system then performs narrative query processing for this pattern and returns the results to the user. If results are found, we call the narrative pattern  \textit{plausible}. 
\end{definition}

\subsection{The Problem of Context-Compatibility}
In this section we investigate the problem of context-compatibility in more detail and discuss suitable solutions how digital libraries can retain contexts in practice.
Contexts define the \textit{scope} in which a piece of information can be fused with other statements. 
This means that a context has to involve all information that need to be known to validate some larger, fused piece of information. 
But unfortunately, essential parts of contexts may get lost during information extraction.  
Generally speaking, problems with context compatibility come in at least two distinct flavors: \textit{constraining contexts and correspondence contexts}. 
Constraining contexts scope the validity of fusions of statements over the entire query, i.e., for some statements in a substitution, a fusion is impossible because they have been extracted from contradicting contexts. 
In contrast, correspondence contexts limit the actual fusion of individual pieces of knowledge between which a fusion would generally be possible but is not warranted by the data from which the information was extracted.   

For a problematic case with \textit{constraining contexts} consider the following example:
\begin{example}
``We report a case of a 62-year-old man who developed cerebral venous sinus thrombosis with subarachnoid hemorrhage and concomitant thrombocytopenia, which occurred 13 days after ChAdOx1 nCov-19 injection.``~\cite{berezne2021deterioration}  

Among others we may extract the following statements:

\begin{itemize}
\item (\textit{patient}, \textit{vaccinated by}, \textit{ChAdOx1 nCov-19})
\item (\textit{patient}, \textit{suffered from}, \textit{cerebral venous sinus thrombosis})
\end{itemize}

But the statement that some patient suffered from cerebral venous sinus thrombosis is only sensible within the context of this particular \textit{patient record}. 
Unfortunately, there is no information whether the statement can be generalized to other patients. 
Thus if the extractions' context (e.g., the patient's age, or that he was recently vaccinated) is lost, information fusions or reasoning processes relying on this specific piece of information may produce invalid results and even run into inconsistencies.
\end{example}

In brief, constructing knowledge bases with insufficiently contextualized statements and then using them to answer complex query patterns may result in invalid answers:
Vaccinations with \textit{ChAdOx1 nCov-19} may indeed lead to a \textit{pneumonia} although probably not in all contexts.

For a problematic case with \textit{corresponding contexts} consider the following example:
\begin{example}
``Secondary analyses found increased risk of CVST after ChAdOx1 nCoV-19 vaccination (4.01, 2.08 to 7.71 at 8-14 days), after BNT162b2 mRNA vaccination (3.58, 1.39 to 9.27 at 15-21 days), and after a positive SARS-CoV-2 test.``~\cite{pmid34446436}

We may extract the following statements:

\begin{itemize}
\item (\textit{ChAdOx1 nCov-19}, \textit{observed condition}, \textit{CVST})
\item (\textit{BNT162 Vaccine}, \textit{observed condition}, \textit{CVST})
\item (\textit{CVST}, \textit{risk after vaccination}, \textit{4.01})
\item (\textit{CVST}, \textit{risk after vaccination}, \textit{3.58})
\end{itemize}

Now information fusion for answering the query (?x, \textit{observed condition}, \textit{CVST}) AND (\textit{CVST}, \textit{risk after vaccination}, ?y). would compute the Cartesian product producing four results (two of which are correct, while the other two are incorrect). This is because the \textit{binary extraction} has lost the information, which risk factor belongs to which vaccine. 
\end{example}

In brief, although all statements are mentioned within the close scope of a clinical trial having inclusion and exclusion criteria, an information extraction process may loose how statements belong together within that context.

Here the text expresses a ternary relation between \textit{vaccines}, \textit{conditions} and \textit{probabilities} that is broken down into binary relations.
Moreover, note that this is not an artifact of automatic processes, as even manual extraction may yield the same result because of the restriction of using only binary relations. 

In conclusion, although all of our example statements were \textit{syntactically} correct, vital \textit{semantics} have been lost because the context was neglected.
This forms a serious threat to the \textbf{validity} of query results, i.e., even correctly extracted but subsequently fused statements may not always produce valid answers in query processing or reasoning. 
Specifically, invalid answers are those cases that do not match the user's context or connect statements that do not belong together.

Since these problems are the main reason we argue to move towards narrative information access, we will take a closer look at possible remedies in the following section.

\subsection{Maintaining Contexts in Digital Libraries}
So how can we retain contexts in practical digital library projects? 
This subsection discusses research and methods to combat both loss of constraining contexts and loss of correspondence contexts.

\paragraph{N-ary Relations.} 
Ernst et al.~\cite{ernst2018highlife} proposed an n-ary extraction method to precisely retain complex relations, e.g., a relation \textit{vaccinated\_patients\_suffer} that involves the \textit{target group}, \textit{vaccine} and \textit{side effects}.
However designing appropriate n-ary relation signatures a-priori is challenging because it requires extensive domain knowledge.
The authors collected examples to train a suitable extraction model for their relations.
In addition, they performed partial reasoning to compose partial statements to n-ary statements because their extraction method was also limited to sentences. 
The reasoning step helped to increase the extraction recall but required the definition of rules (which facts should be composed).
Although n-ary relations are strongly appreciated, practical extraction methods hardly support them because defining signatures, providing enough training examples, and formulating reasoning constraints is an exhausting task.

\paragraph{Explicit Context Models.}
McCarthy introduced an explicit context model based on the first-order predicate logic~\cite{mccarthy93contexts}.
The model allows users to formulate context conditions for arbitrary statements explicitly. 
In addition, he discussed relations between contexts, e.g., one context might \textit{specialize} another context.
Hand-crafted rules were then formulated to determine how to combine contexts and their enclosed statements.
\textit{VIKEF} is an example digital library project supporting explicit context information in an RDF knowledge base~\cite{stoermer2006dlcontextualizedknowledgebase}.

\paragraph{Implicit Contexts.}
We proposed using document references as an implicit and practical context model~\cite{kroll2020tpdl}.
We suggested to store references to the source documents when harvesting statements from it. 
These references were then used to estimate which statements can safely be combined to produce valid answers. 
When combining only statements extracted from the same document, the resulting precision in a downstream application will increase, but the recall is bound to decrease.
We therefore proposed measures to estimate \textit{compatibility} between contexts to flexibly manage the precision/recall trade-off, e.g., text and author similarities.

Such implicit context models might be suitable candidates to retain context in digital libraries because they are cheap to maintain, i.e., only references to the statements' sources must be retained.
But their quality and explainability are somewhat limited, e.g., how should we explain why two documents are context-compatible based on some text similarity measure.
Keyword extraction might be a good method to retrieve context proxies here; See YAKE~\cite{campos2020yake} for example.
In summary, implicit context models are easy to use and may yield good precision, but estimating context-compatibility remains challenging, and the overall quality achieved might still not be good enough for digital libraries.

\paragraph{Provenance.} Provenance information is often understood to be any kind of information that may validate some statement's quality or origin~\cite{ProvenanceInLinkedData}. 
Provenance might range from storing a reference to the statement's origin to storing information about the creation process, e.g., author, release date, point in time, and more.
The Prov-O Ontology Description is a common standard for defining and storing general provenance information~\cite{ProvO}. 
Prov-O supports complex provenance graphs to describe the origin of some statements.
As an alternative, the Wikidata project supports qualifiers (property-value pairs) to retain provenance for its statements~\cite{vrandevcic2014wikidata}, e.g., references, determination methods, time and location information.

Nevertheless, using qualifiers and provenance information in practical applications, especially in query processing, remains an exception. 
Returning to our example, how could we use a qualifier information about the \textit{62-year-old man} in query processing?
Should we formulate hand-crafted rules on how different provenance information affects the actual query processing? 
How do we know when qualifiers describe the same or a compatible context? 
We understand Prov-O and provenance in general as possible implementations to store contexts.
However they do not provide a ready-to-use solution to retain both by default.
Domain experts and digital library curators must carefully define corresponding statements and describe how they are used for a practical application.

\section{Narrative Query Processing in Practice -- Case Studies}
We performed case studies to understand the benefits and limitations of narrative information access.
In particular, we built on our publicly available narrative retrieval system called \textit{Narrative Query Graphs for Entity-Interaction Document Retrieval} by \cite{kroll2021narrativequerygraphs}. 
We built a working document retrieval system that allows formulating information needs as graph patterns, i.e.,  entities and their corresponding interactions.
We transformed biomedical document abstracts into a graph representation called document graph as knowledge bases.
Then the retrieval system allows matching user queries against these document graphs and returns all matches.
Since document graphs match queries only within single documents, contexts are to some degree considered in query processing because the context can quite safely be assumed to be consistent within each document abstract.

\subsection{Narrative Query Graphs for Covid 19}
In cooperation with pharmaceutical domain experts, the Robert-Koch Institute in Germany and the ZB MED library, 
we enhanced the narrative retrieval system to answer Covid 19-related research questions: 
\begin{enumerate}
\item We included the LitCovid collection from PubMed (peer-reviewed articles about Covid 19) and the latest Covid 19-related pre-prints supplied by ZB MED~\cite{langnickel2021covid,langnickel2021covidpreview}. These pre-prints can be accessed via their Preview service\footnote{\url{https://preview.zbmed.de/}}.
\item We developed a vaccine entity vocabulary by utilizing Wikidata and the Medical Subject Headings (MeSH). In addition, we derived an entity for Long Covid 19 from MeSH.
\end{enumerate}

The prototype of the enhanced narrative query system is publicly available\footnote{ \url{http://www.pubpharm.de/services/prototypes/narratives/}}.
In the following we investigate whether typical research questions from the pharmacy domain can be translated into narrative query graphs and how helpful such searches are in practice.
Please note that this case study does not yet contain a comprehensive evaluation. 
We are currently preparing a large-scale study with our partners.

\paragraph{Long Covid Related Questions} 
The development of the Covid 19 pandemics has shown that \textit{Long Covid} is a severe threat to a patient's health.
So what are \textit{common symptoms that are reported for Long Covid}?
We formulated the following query graph: (\textit{post-acute COVID-19 syndrome}, \textit{associated}, \textit{?X(Disease)}). 
\textit{?X(Disease)} means that we search with a variable named \textit{?X} that should be substituted by entities of the type \textit{Disease}.
\textit{Post-acute COVID-19 syndrome} is an entity from the Medical Subject Headings (MeSH)\footnote{\url{https://meshb.nlm.nih.gov/record/ui?ui=C000711409}}.
The system responded with a list of commonly known conditions such as \textit{Fatigue (44)}, \textit{Dyspnea (19)}, \textit{Anossmia (10)}, \textit{Cognitive Dysfunction (9)} and \textit{Headache (7)}. 
The number in brackets refers to how many documents share the corresponding variable substitution.
The system can show the origin of the extraction, i.e., the sentence in which the pattern was matched.
However also substitutions such as \textit{Covid 19 (143)} and \textit{Infections (61)} were not helpful.

We adjusted the previous query to search for patient cases: (\textit{post-acute COVID-19 syndrome}, \textit{associated}, \textit{Human}) AND (\textit{Human}, \textit{associated}, \textit{?X(Disease)}) .
Here \textit{Humans} is an entity that stand for \textit{patients}, \textit{men}, \textit{women}, etc. 
The current version of the system did not support searching for specific target groups.
This query could be matched against abstracts such as: \textit{``[...] post-COVID-19 syndrome in patients with primary Sjogren's syndrome (pSS) affected by acute SARS-CoV-2 infection. [...] More than 40\% of pSS patients reported the persistence of four symptoms or more, including anxiety/depression (59\%), arthralgias (56\%), sleep disorder (44\%), fatigue (40\%), anosmia (34\%) and myalgias (32\%).``}~\cite{pmid34874821}
Here the implicit context ensured that both statements must be matched against a single abstract.
But the number of found results were decreased: \textit{Fatigue (15)}, \textit{Dyspnea (8)}, \textit{Cognitive Dysfunction (4)} and \textit{Headache (3)}. 

A quick look over both results revealed that publications were missed because they did not explicitly contain the entity \textit{post-acute COVID-19 syndrome}. 
Instead, publications may describe Covid 19 infections and observations made six months later.
Here entity linking did not detect the explicit entity.

\paragraph{Vaccinations.} 
We formulated a query to list commonly used vaccines that are associated with Covid 19: (\textit{Covid 19}, \textit{associated}, \textit{?X(Vaccine)}. 
Helpful substitutions were for example: \textit{BTN162 aka Pfizer (175)}, \textit{ChAdOx1 nCoV-19 aka Astra Zeneca (79)}, and \textit{2019-nCoV Vaccine mRNA-1273 aka Moderna (76)}.
In addition, miss leading substitutions like \textit{Vaccine (3472)} and \textit{Covid-19 Vaccines (685)} were found and not helpful because they were far too general.
We enhanced the query by asking for common side effects of \textit{ChAdOx1 nCoV-19}: (\textit{ChAdOx1 nCoV-19}, \textit{associated}, \textit{?X(Disease)}.
Substitutions such as \textit{Thrombosis (93)}, \textit{Thrombocytopenia (79)}, and \textit{CVST (18)} were found.
The system yielded also not helpful results like \textit{Covid-19 (79)} and \textit{Infections (27)} caused by wrong extractions.
Again, we added the \textit{Human} entity to precisely query for studies: (\textit{Human}, \textit{associated}, \textit{?X(Disease)}) AND (\textit{Chadox1 Ncov-19}, \textit{associated}, \textit{Human}). 
Here we could quickly find a case study~\cite{pmid34729931} for CVST investigation.

\paragraph{Treatments.}
We were also interested in queries that consider treatments for Covid-19 symptoms. 
Therefore, we formulated the query: (\textit{?X(Drug)}, \textit{treats}, \textit{Covid 19}).
Helpful substitutions were \textit{Hydroxychloroquiene (829)} and \textit{Remdesivir (581)}.
The system's provenance information (matched sentences) showed that the system found the statement in sentences like: \textit{``An example of which is remdesivir which has now been approved for use in COVID-19 patients by the US Food and Drug Administration.``}~\cite{pmid34956606}
We rewrote the query by integrating the patient again, similar to the previous approaches.
Here we retrieved matches such as \textit{``We identified 55 patients who were treated with remdesivir for COVID-19 and analyzed inflammatory markers and clinical outcomes.``}~\cite{pmid34406670}

\paragraph{Discussion.}
The case study showed that narrative information access indeed could support typical tasks like generating structured overviews of the latest literature or quickly finding precise hits:
On the one hand, suitable substitutions for \textit{Long Covid 19 symptoms} or \textit{Covid 19 drug treatments} were indeed found, thus successfully structuring the latest literature. 
On the other hand, the expressive query format enabled the integration of \textit{patients} in the query to ensure that the results had to connect the disease or drug to a concrete target group.

As a small caveat, note that all queries were matched only against implicit document contexts, ensuring the statements' context compatibility.
In this way retaining the context for query processing came cheap: 
The origin of the statements needed to be stored and the query processing had to be restricted to document graphs.
Of course, this (overly careful) restriction to document graphs also comes with severe limitations since combining knowledge from different sources is a common practice and vital necessity in scientific research. 
While the precision in our query tasks was very high and thus matches were accurate, the respective recall was admittedly marginal. 
More open yet effective measures for controlling context-compatibility than using documents graphs will be needed to build large-scale practical narrative retrieval systems (as previously discussed in section 2.2).

\subsection{Narrative Query Graphs in Political Sciences}
In cooperation with the specialized information service for political sciences ~\cite{SchardelmannOtto2018pollux}(Pollux)\footnote{\url{https://www.pollux-fid.de}} we were interested how the political sciences can benefit from narrative information access.
We asked an expert (Ph.D. in political sciences) to study the biomedical narrative query graph retrieval system. 
He then formulated questions that would be of interest in political sciences. 
Due to the lack of available knowledge bases we could not realize a practical retrieval system here. 
Instead, we went through two of his questions and argue in the following how they could be answered and why narrative information access is vital.
In addition we report on opportunities and potential obstacles in political sciences.
In the following we picked two of his questions as showcases:

\begin{enumerate}
    \item \textit{How do heads of government in Latin America and Scandinavia present the question what action is needed in relation to climate change?}
    \item \textit{How do Germany's major daily newspapers negotiate the course of the refugee crisis in 2015 and 2016?}
\end{enumerate}

So why do we need narrative information access to answer his questions?
The main reason here is that both questions asked to combine several information: 
For the first question, we have to combine statements about climate change in the time period of corresponding presidents (temporal and location context).
The temporal and location contexts and the source of information (the heads of government) are vital to determine statements' validity.
For the second question, we have to generate a structured overview of statements and viewpoints (e.g., conservative, progressive, etc.) from daily newspapers about the refugee crisis in 2015 and 2016 (temporal context, framing, and wording).
The selection of keywords (wording) may express different viewpoints.
Again, context (e.g., the kind and target group of a newspaper) was vital to align the statements with a certain viewpoint. 

Parts of both queries could be answered with today's knowledge bases already.
Consider, for example, the usage of Wikidata:
Concerning question (1), formulating a SPARQL query allowed us to retrieve a list of heads of governments in both geographical regions. 
And we could also combine the results with their temporal context:
\begin{itemize}
\item (\textit{?country}, \textit{head\_of\_state}, \textit{?stmt}) AND
\\ (\textit{?stmt}, \textit{head\_of\_state}, \textit{?person}) AND
\\ (\textit{?stmt}, \textit{start\_time}, \textit{?time}) AND
\\ (\textit{?country}, \textit{part\_of}, \textit{Latin America}).
\end{itemize}

Note, the \textit{?stmt} notation is necessary to query Wikidata for qualifiers.
This query yielded 66 results.

Concerning question (2), major newspaper could be easily identified by querying Wikidata: (\textit{?newspaper}, \textit{instance\_of}, \textit{daily newspaper}) AND (\textit{?newspaper}, \textit{country}, \textit{Germany}). 
Querying Wikidata resulted in 58 newspapers.
Newspapers are often associated with a \textit{political ideology}.
And indeed, Wikidata stores information that the \textit{Frankfurter Allgemeine Zeitung (FAZ)} has the \textit{political ideology} \textit{liberal conservatism}\footnote{\url{https://www.wikidata.org/wiki/Q10184}}.
In this way we could derive additional context information when analyzing statements from a newspaper.
Note that this might be a good approximation but newspapers might also include articles that follow different ideologies.

The next part would include context-sensitive information retrieval based on the Wikidata results.
To answer both questions, we had to rely on texts, e.g., from Pollux or specialized knowledge bases for claims such as  ClaimsKG~\cite{tchechmedjiev2019claimskg}.
Here a comprehensive extraction is necessary to identify statements in texts.

But even if a knowledge base had been available, question (2) asked for different levels of granularity regarding the context of statements. 
In a simple scenario, it might be enough to extract statements from news articles and cluster them by their \textit{political ideology} from Wikidata if available.
However guest commentary or changes in the editorial board might include statements that stemmed from a different ideology. 
Therefore, we have to classify the ideology based on an article's wording and framing, and may not solely rely on the general ideology of a newspaper.

\paragraph{Challenges.}
Political sciences have a broad range of essential concepts, e.g., viewpoints, schools of thought, and ambiguous terms.
These concepts are hard to identify in a text, unlike biomedical entities.
Here wording and framing of texts might determine the viewpoint, whereas a drug in medicine remains the same drug regardless of wording.
Moreover, central terms like "Democracy" or "Society" are not unambiguously defined and can be interpreted differently, depending on a school of thought.
Furthermore, even if we identify the concepts, extracting structured information remains challenging.
Statements in this domain are more complex than just expressing a binary relation between a patient and a disease condition.

\balance 
These issues have to be addressed to realize a convenient narrative information access.
Although solving them remains challenging, the previous cases showed that political sciences could benefit from such access.
Structuring publications into schools of thought or clustering viewpoints regarding a topic would be beneficial here.
Moreover, without considering the context of information, such access could hardly be realized.

\subsection{Investigating Common Knowledge Bases}
After we performed both case studies, we also were interested in the generalizability of the benefits of narrative information access to other domains.
We first had a look at publicly available knowledge bases for their application and possible issues.

Interestingly, the following statement is included in Wikidata\footnote{\url{https://www.wikidata.org/wiki/Q76} 01.2022}:

\begin{itemize}
\item (\textit{Barack Obama}, \textit{born in}, \textit{Kenya})
\end{itemize}

In Wikidata this statement is complemented by a qualifier that states \textit{mentioned in a conspiracy theory}. 
A qualifier is a statement about some other statement, i.e., a property-value pair attached to a statement.
But this incorrect statement that \textit{Barack Obama was born in Kenya} can only be sensible when considering it in the context of some \textit{conspiracy theory}. 
Wikidata marks this data in their user interface by an colour encoding: \textit{green} for fact-checked and \textit{red} for not fact-checked.
However the decision whether something is fact-checked or not is often not easy, e.g., partially fact-checked statements.
In addition, different school of thoughts may accept or reject a certain statement.
And having a general decision here, whether something is \textit{true} or not, remains open.

We found another interesting example in the real-world knowledge base DBpedia\footnote{\url{https://dbpedia.org/page/Barack_Obama} 01.2022}.
\begin{itemize}
    \item (\textit{Barack Obama}, \textit{was}, \textit{Senator of Illinois})
    \item (\textit{Barack Obama}, \textit{predecessor}, \textit{Peter G. Fitzgerald})
    \item (\textit{Barack Obama}, \textit{was}, \textit{U.S. President})
    \item (\textit{Barack Obama}, \textit{predecessor}, \textit{George W. Bush})
\end{itemize}

Suppose a user asks the following query: \textit{Who was the predecessor of the U.S. President Barack Obama?}
In that case the results are \textit{George W. Bush} (correct) and \textit{Peter G. Fitzgerald} (wrong).
Thus querying DBpedia with such queries can lead to wrong results.
The example query could have been answered correctly if the connection between the statements had been retained.

Both examples show that the loss of context is also an issue in common knowledge bases. 
Information can quickly be broken down lossy and cannot be reassembled lossless afterward.

\section{Discussion}
Narrative information access ensures that the binding process must consider contexts when making a narrative plausible.
Here bindings must be context-compatible which ensures that the bindings form valid answers.
We do not claim that knowledge bases cannot do the job.
But if they are built without considering context and statements are restricted to triples, then information is broken down in a lossy fashion and cannot be reassembled lossless afterward.
Thus contexts definitely have to be considered when designing knowledge bases to supply narrative information access.

\subsection{Generalizability}
Although we made our central use case in the biomedical domain, we argue that we can generalize our findings across domains.
The Obama examples show how easily context can be lost in common knowledge bases.
In addition, we reported on opportunities and challenges in political sciences.
Here proposed use cases showed how beneficial narrative information access could be.
Due to the lack of structured knowledge bases, we could hardly realize an access here.
But context like temporal periods or a newspaper's viewpoint is essential to answer narrative queries correctly.

\subsection{Benefits for Digital Libraries}
The Covid 19 pandemics has shown how important it is to carefully handle scientific claims.
Tearing such claims apart from the original lines of arguments has caused many miss leading debates (based on fake news) and movements across the world.
Digital libraries should head for a more comprehensive knowledge curation by allowing narrative information access.
Here the vital contexts are considered when answering queries.
Our case study has shown how context-aware query systems can be applied to Covid 19 related questions.
Although our study lacked a comprehensive evaluation, we demonstrate such benefits in practice:
Narrative Information access allows to structure the latest literature or quickly find suitable information.
Realizing and implementing suitable workflows may be cost-intensive, but digital libraries can benefit from them.

\subsection{Future Work} 
A new challenge that has to be addressed for narrative information access is the growing heterogeneity of data sources with digital libraries, such as textual sources, image collections, experimental data or structured knowledge bases.
Research data sets are a good consideration to link narrative queries against~\cite{nagel2021datasetlinking}.
Making these heterogeneous repositories accessible in a unified way and integrating their different kinds of information requires effective access paths that often have to be intelligently customized to the content types. 
For narrative information access this means that bindings on (sub-)graphs of narrative queries have to be computed against extractions (either precomputed or extracted on-the-fly) from different media.  
Investigating such extraction is thus essential for broader applicability of narrative retrieval systems.

\section{Conclusion}
Although knowledge bases allow effective access paths in digital libraries, we demonstrated their limitations when handling narrative information.
Here information, originally stated in coherent lines of arguments, can be broken into pieces that cannot be reassembled lossless afterward.
This paper defines narrative information access as an extension to common knowledge base querying.
Here the context of statements must be retained and considered to produce valid answers when querying narrative information.
Realizing narrative information access in digital libraries can be cost-intensive in practice, but like the case study for Covid 19 retrieval has shown, implicit document contexts may approximate it.
The examples of Barack Obama in common knowledge bases, our investigation in Covid 19 related questions, and the discussion in political sciences have shown how beneficial narrative information access can be.
Even now existing methods and techniques can be used to implement narrative information access in digital libraries reliably.
However handling heterogeneous library content (research data, tables, images, etc.) would be the next step to enhance such access further.

\section*{Acknowledgment}
Supported by the Deutsche Forschungsgemeinschaft (DFG, German Research Foundation): PubPharm – the Specialized Information Service for Pharmacy (Gepris 267140244).

\bibliographystyle{ACM-Reference-Format}
\bibliography{references}


\begin{thebibliography}{28}


\ifx \showCODEN    \undefined \def \showCODEN     #1{\unskip}     \fi
\ifx \showDOI      \undefined \def \showDOI       #1{#1}\fi
\ifx \showISBNx    \undefined \def \showISBNx     #1{\unskip}     \fi
\ifx \showISBNxiii \undefined \def \showISBNxiii  #1{\unskip}     \fi
\ifx \showISSN     \undefined \def \showISSN      #1{\unskip}     \fi
\ifx \showLCCN     \undefined \def \showLCCN      #1{\unskip}     \fi
\ifx \shownote     \undefined \def \shownote      #1{#1}          \fi
\ifx \showarticletitle \undefined \def \showarticletitle #1{#1}   \fi
\ifx \showURL      \undefined \def \showURL       {\relax}        \fi
\providecommand\bibfield[2]{#2}
\providecommand\bibinfo[2]{#2}
\providecommand\natexlab[1]{#1}
\providecommand\showeprint[2][]{arXiv:#2}

\bibitem[\protect\citeauthoryear{Auer, Bizer, Kobilarov, Lehmann, Cyganiak, and
  Ives}{Auer et~al\mbox{.}}{2007}]%
        {auer2007dbpedia}
\bibfield{author}{\bibinfo{person}{S{\"o}ren Auer}, \bibinfo{person}{Christian
  Bizer}, \bibinfo{person}{Georgi Kobilarov}, \bibinfo{person}{Jens Lehmann},
  \bibinfo{person}{Richard Cyganiak}, {and} \bibinfo{person}{Zachary Ives}.}
  \bibinfo{year}{2007}\natexlab{}.
\newblock \showarticletitle{Dbpedia: A nucleus for a web of open data}.
\newblock In \bibinfo{booktitle}{\emph{The semantic web}}.
  \bibinfo{publisher}{Springer}, \bibinfo{address}{Busan, Korea},
  \bibinfo{pages}{722--735}.
\newblock


\bibitem[\protect\citeauthoryear{B{\'e}rezn{\'e}, Bougon, Blanc-Jouvan,
  Gendron, Janssen, Muller, Bertil, Desvard, Presot, Terrier,
  et~al\mbox{.}}{B{\'e}rezn{\'e} et~al\mbox{.}}{2021}]%
        {berezne2021deterioration}
\bibfield{author}{\bibinfo{person}{Alice B{\'e}rezn{\'e}},
  \bibinfo{person}{David Bougon}, \bibinfo{person}{Florence Blanc-Jouvan},
  \bibinfo{person}{Nicolas Gendron}, \bibinfo{person}{Cecile Janssen},
  \bibinfo{person}{Michel Muller}, \bibinfo{person}{S{\'e}bastien Bertil},
  \bibinfo{person}{Florence Desvard}, \bibinfo{person}{Isabelle Presot},
  \bibinfo{person}{Benjamin Terrier}, {et~al\mbox{.}}}
  \bibinfo{year}{2021}\natexlab{}.
\newblock \showarticletitle{Deterioration of vaccine-induced immune thrombotic
  thrombocytopenia treated by heparin and platelet transfusion: Insight from
  functional cytometry and serotonin release assay}.
\newblock \bibinfo{journal}{\emph{Research and Practice in Thrombosis and
  Haemostasis}} \bibinfo{volume}{5}, \bibinfo{number}{6}
  (\bibinfo{year}{2021}), \bibinfo{pages}{e12572}.
\newblock


\bibitem[\protect\citeauthoryear{Brito-Zerón, Acar-Denizli, Romão, Armagan,
  Seror, Carubbi, Melchor, Priori, Valim, Retamozo, Pasoto, Trevisani, Hofauer,
  Szántó, Inanc, Hernández-Molina, Sebastian, Bartoloni, Devauchelle-Pensec,
  Akasbi, Giardina, Bandeira, Sisó-Almirall, and Ramos-Casals}{Brito-Zerón
  et~al\mbox{.}}{2021}]%
        {pmid34874821}
\bibfield{author}{\bibinfo{person}{P. Brito-Zerón}, \bibinfo{person}{N.
  Acar-Denizli}, \bibinfo{person}{V.~C. Romão}, \bibinfo{person}{B. Armagan},
  \bibinfo{person}{R. Seror}, \bibinfo{person}{F. Carubbi}, \bibinfo{person}{S.
  Melchor}, \bibinfo{person}{R. Priori}, \bibinfo{person}{V. Valim},
  \bibinfo{person}{S. Retamozo}, \bibinfo{person}{S.~G. Pasoto},
  \bibinfo{person}{V.~F.~M. Trevisani}, \bibinfo{person}{B. Hofauer},
  \bibinfo{person}{A. Szántó}, \bibinfo{person}{N. Inanc},
  \bibinfo{person}{G. Hernández-Molina}, \bibinfo{person}{A. Sebastian},
  \bibinfo{person}{E. Bartoloni}, \bibinfo{person}{V. Devauchelle-Pensec},
  \bibinfo{person}{M. Akasbi}, \bibinfo{person}{F. Giardina},
  \bibinfo{person}{M. Bandeira}, \bibinfo{person}{A. Sisó-Almirall}, {and}
  \bibinfo{person}{M. Ramos-Casals}.} \bibinfo{year}{2021}\natexlab{}.
\newblock \showarticletitle{{{P}ost-{C}{O}{V}{I}{D}-19 syndrome in patients
  with primary {S}jögren's syndrome after acute {S}{A}{R}{S}-{C}o{V}-2
  infection}}.
\newblock \bibinfo{journal}{\emph{Clin Exp Rheumatol}} \bibinfo{volume}{39
  Suppl 133}, \bibinfo{number}{6} (\bibinfo{year}{2021}),
  \bibinfo{pages}{57--65}.
\newblock


\bibitem[\protect\citeauthoryear{Butnariu, Look, Grillo, Tabish, McGarvey, and
  Pranjol}{Butnariu et~al\mbox{.}}{2022}]%
        {pmid34956606}
\bibfield{author}{\bibinfo{person}{A.~B. Butnariu}, \bibinfo{person}{A. Look},
  \bibinfo{person}{M. Grillo}, \bibinfo{person}{T.~A. Tabish},
  \bibinfo{person}{M.~J. McGarvey}, {and} \bibinfo{person}{M.~Z.~I. Pranjol}.}
  \bibinfo{year}{2022}\natexlab{}.
\newblock \showarticletitle{{{S}{A}{R}{S}-{C}o{V}-2-host cell surface
  interactions and potential antiviral therapies}}.
\newblock \bibinfo{journal}{\emph{Interface Focus}} \bibinfo{volume}{12},
  \bibinfo{number}{1} (\bibinfo{date}{Feb} \bibinfo{year}{2022}),
  \bibinfo{pages}{20200081}.
\newblock


\bibitem[\protect\citeauthoryear{Campos, Mangaravite, Pasquali, Jorge, Nunes,
  and Jatowt}{Campos et~al\mbox{.}}{2020}]%
        {campos2020yake}
\bibfield{author}{\bibinfo{person}{Ricardo Campos}, \bibinfo{person}{Vítor
  Mangaravite}, \bibinfo{person}{Arian Pasquali}, \bibinfo{person}{Alípio
  Jorge}, \bibinfo{person}{Célia Nunes}, {and} \bibinfo{person}{Adam Jatowt}.}
  \bibinfo{year}{2020}\natexlab{}.
\newblock \showarticletitle{YAKE! Keyword extraction from single documents
  using multiple local features}.
\newblock \bibinfo{journal}{\emph{Information Sciences}}  \bibinfo{volume}{509}
  (\bibinfo{year}{2020}), \bibinfo{pages}{257--289}.
\newblock
\showISSN{0020-0255}
\urldef\tempurl%
\url{https://doi.org/10.1016/j.ins.2019.09.013}
\showDOI{\tempurl}


\bibitem[\protect\citeauthoryear{Ernst, Siu, and Weikum}{Ernst
  et~al\mbox{.}}{2018}]%
        {ernst2018highlife}
\bibfield{author}{\bibinfo{person}{Patrick Ernst}, \bibinfo{person}{Amy Siu},
  {and} \bibinfo{person}{Gerhard Weikum}.} \bibinfo{year}{2018}\natexlab{}.
\newblock \showarticletitle{HighLife: Higher-Arity Fact Harvesting}. In
  \bibinfo{booktitle}{\emph{Proceedings of the 2018 World Wide Web Conference}}
  (Lyon, France) \emph{(\bibinfo{series}{WWW '18})}.
  \bibinfo{publisher}{International World Wide Web Conferences Steering
  Committee}, \bibinfo{address}{Republic and Canton of Geneva, CHE},
  \bibinfo{pages}{1013–1022}.
\newblock
\showISBNx{9781450356398}
\urldef\tempurl%
\url{https://doi.org/10.1145/3178876.3186000}
\showDOI{\tempurl}


\bibitem[\protect\citeauthoryear{Freeman}{Freeman}{2011}]%
        {freeman2011argument}
\bibfield{author}{\bibinfo{person}{James~B Freeman}.}
  \bibinfo{year}{2011}\natexlab{}.
\newblock \bibinfo{booktitle}{\emph{Argument Structure:: Representation and
  Theory}}. Vol.~\bibinfo{volume}{18}.
\newblock \bibinfo{publisher}{Springer Science \& Business Media},
  \bibinfo{address}{Berlin/Heidelberg, Germany}.
\newblock


\bibitem[\protect\citeauthoryear{Kilicoglu, Shin, Fiszman, Rosemblat, and
  Rindflesch}{Kilicoglu et~al\mbox{.}}{2012}]%
        {kilicoglu2012semmeddb}
\bibfield{author}{\bibinfo{person}{Halil Kilicoglu}, \bibinfo{person}{Dongwook
  Shin}, \bibinfo{person}{Marcelo Fiszman}, \bibinfo{person}{Graciela
  Rosemblat}, {and} \bibinfo{person}{Thomas~C. Rindflesch}.}
  \bibinfo{year}{2012}\natexlab{}.
\newblock \showarticletitle{{SemMedDB: a PubMed-scale repository of biomedical
  semantic predications}}.
\newblock \bibinfo{journal}{\emph{Bioinformatics}} \bibinfo{volume}{28},
  \bibinfo{number}{23} (\bibinfo{date}{10} \bibinfo{year}{2012}),
  \bibinfo{pages}{3158--3160}.
\newblock
\showISSN{1367-4803}
\urldef\tempurl%
\url{https://doi.org/10.1093/bioinformatics/bts591}
\showDOI{\tempurl}


\bibitem[\protect\citeauthoryear{Klungel and Pottegård}{Klungel and
  Pottegård}{2021}]%
        {pmid34446436}
\bibfield{author}{\bibinfo{person}{O.~H. Klungel} {and} \bibinfo{person}{A.
  Pottegård}.} \bibinfo{year}{2021}\natexlab{}.
\newblock \showarticletitle{{{S}trengthening international surveillance of
  vaccine safety}}.
\newblock \bibinfo{journal}{\emph{BMJ}}  \bibinfo{volume}{374}
  (\bibinfo{date}{08} \bibinfo{year}{2021}), \bibinfo{pages}{n1994}.
\newblock


\bibitem[\protect\citeauthoryear{Kroll, Kalo, Nagel, Mennicke, and Balke}{Kroll
  et~al\mbox{.}}{2020a}]%
        {kroll2020tpdl}
\bibfield{author}{\bibinfo{person}{Hermann Kroll},
  \bibinfo{person}{Jan-Christoph Kalo}, \bibinfo{person}{Denis Nagel},
  \bibinfo{person}{Stephan Mennicke}, {and} \bibinfo{person}{Wolf-Tilo Balke}.}
  \bibinfo{year}{2020}\natexlab{a}.
\newblock \showarticletitle{Context-Compatible Information Fusion for
  Scientific Knowledge Graphs}. In \bibinfo{booktitle}{\emph{Digital Libraries
  for Open Knowledge}}. \bibinfo{publisher}{Springer}, \bibinfo{address}{Lyon,
  France}, \bibinfo{pages}{33--47}.
\newblock
\showISBNx{978-3-030-54956-5}
\urldef\tempurl%
\url{https://doi.org/10.1007/978-3-030-54956-5_3}
\showDOI{\tempurl}


\bibitem[\protect\citeauthoryear{Kroll, Nagel, and Balke}{Kroll
  et~al\mbox{.}}{2020b}]%
        {kroll2020er}
\bibfield{author}{\bibinfo{person}{Hermann Kroll}, \bibinfo{person}{Denis
  Nagel}, {and} \bibinfo{person}{Wolf-Tilo Balke}.}
  \bibinfo{year}{2020}\natexlab{b}.
\newblock \showarticletitle{Modeling Narrative Structures in Logical Overlays
  on Top of Knowledge Repositories}. In \bibinfo{booktitle}{\emph{Conceptual
  Modeling}}. \bibinfo{publisher}{Springer}, \bibinfo{address}{Vienna,
  Austria}, \bibinfo{pages}{250--260}.
\newblock
\showISBNx{978-3-030-62522-1}
\urldef\tempurl%
\url{https://doi.org/10.1007/978-3-030-62522-1_18}
\showDOI{\tempurl}


\bibitem[\protect\citeauthoryear{Kroll, Nagel, Kunz, and Balke}{Kroll
  et~al\mbox{.}}{2021a}]%
        {kroll2021narrativebindings}
\bibfield{author}{\bibinfo{person}{Hermann Kroll}, \bibinfo{person}{Denis
  Nagel}, \bibinfo{person}{Morris Kunz}, {and} \bibinfo{person}{Wolf-Tilo
  Balke}.} \bibinfo{year}{2021}\natexlab{a}.
\newblock \showarticletitle{Demonstrating Narrative Bindings: Linking
  Discourses to Knowledge Repositories}. In \bibinfo{booktitle}{\emph{Fourth
  Workshop on Narrative Extraction From Texts, Text2Story@ECIR2021}}
  \emph{(\bibinfo{series}{{CEUR} Workshop Proceedings},
  Vol.~\bibinfo{volume}{2860})}. \bibinfo{publisher}{CEUR-WS.org},
  \bibinfo{pages}{57--63}.
\newblock
\urldef\tempurl%
\url{http://ceur-ws.org/Vol-2860/paper7.pdf}
\showURL{%
\tempurl}


\bibitem[\protect\citeauthoryear{Kroll, Pirklbauer, Kalo, Kunz, Ruthmann, and
  Balke}{Kroll et~al\mbox{.}}{2021b}]%
        {kroll2021narrativequerygraphs}
\bibfield{author}{\bibinfo{person}{Hermann Kroll}, \bibinfo{person}{Jan
  Pirklbauer}, \bibinfo{person}{Jan{-}Christoph Kalo}, \bibinfo{person}{Morris
  Kunz}, \bibinfo{person}{Johannes Ruthmann}, {and}
  \bibinfo{person}{Wolf{-}Tilo Balke}.} \bibinfo{year}{2021}\natexlab{b}.
\newblock \showarticletitle{Narrative Query Graphs for Entity-Interaction-Aware
  Document Retrieval}. In \bibinfo{booktitle}{\emph{Towards Open and
  Trustworthy Digital Societies - 23rd International Conference on Asia-Pacific
  Digital Libraries, {ICADL} 2021, Virtual Event, December 1-3, 2021,
  Proceedings}} \emph{(\bibinfo{series}{Lecture Notes in Computer Science},
  Vol.~\bibinfo{volume}{13133})}. \bibinfo{publisher}{Springer},
  \bibinfo{address}{Online}, \bibinfo{pages}{80--95}.
\newblock
\urldef\tempurl%
\url{https://doi.org/10.1007/978-3-030-91669-5\_7}
\showDOI{\tempurl}


\bibitem[\protect\citeauthoryear{Langnickel, Baum, Darms, Madan, and
  Fluck}{Langnickel et~al\mbox{.}}{2021a}]%
        {langnickel2021covid}
\bibfield{author}{\bibinfo{person}{Lisa Langnickel}, \bibinfo{person}{Roman
  Baum}, \bibinfo{person}{Johannes Darms}, \bibinfo{person}{Sumit Madan}, {and}
  \bibinfo{person}{Juliane Fluck}.} \bibinfo{year}{2021}\natexlab{a}.
\newblock \showarticletitle{COVID-19 preVIEW: Semantic Search to Explore
  COVID-19 Research Preprints}.
\newblock In \bibinfo{booktitle}{\emph{Public Health and Informatics}}.
  \bibinfo{publisher}{IOS Press}, \bibinfo{address}{Amsterdam, the
  Netherlands}, \bibinfo{pages}{78--82}.
\newblock
\urldef\tempurl%
\url{https://doi.org/10.3233/SHTI210124}
\showDOI{\tempurl}


\bibitem[\protect\citeauthoryear{Langnickel, Darms, Baum, and Fluck}{Langnickel
  et~al\mbox{.}}{2021b}]%
        {langnickel2021covidpreview}
\bibfield{author}{\bibinfo{person}{Lisa Langnickel}, \bibinfo{person}{Johannes
  Darms}, \bibinfo{person}{Roman Baum}, {and} \bibinfo{person}{Juliane Fluck}.}
  \bibinfo{year}{2021}\natexlab{b}.
\newblock \showarticletitle{preVIEW: from a fast prototype towards a
  sustainable semantic search system for central access to COVID-19 preprints}.
\newblock \bibinfo{journal}{\emph{Journal of EAHIL}} \bibinfo{volume}{17},
  \bibinfo{number}{3} (\bibinfo{date}{Sep.} \bibinfo{year}{2021}),
  \bibinfo{pages}{8--14}.
\newblock
\urldef\tempurl%
\url{https://doi.org/10.32384/jeahil17484}
\showDOI{\tempurl}


\bibitem[\protect\citeauthoryear{L{\'a}szl{\'o}}{L{\'a}szl{\'o}}{2008}]%
        {laszlo2008science}
\bibfield{author}{\bibinfo{person}{J{\'a}nos L{\'a}szl{\'o}}.}
  \bibinfo{year}{2008}\natexlab{}.
\newblock \bibinfo{booktitle}{\emph{The science of stories: An introduction to
  narrative psychology}}.
\newblock \bibinfo{publisher}{Routledge}, \bibinfo{address}{Oxfordshire,
  England, UK}.
\newblock


\bibitem[\protect\citeauthoryear{Lebo, Sahoo, and McGuinness}{Lebo
  et~al\mbox{.}}{2013}]%
        {ProvO}
\bibfield{author}{\bibinfo{person}{T. Lebo}, \bibinfo{person}{S. Sahoo}, {and}
  \bibinfo{person}{D. McGuinness}.} \bibinfo{year}{2013}\natexlab{}.
\newblock \bibinfo{title}{{PROV-O: The PROV Ontology}}.
\newblock \bibinfo{howpublished}{\url{https://www.w3.org/TR/prov-o/}}.
\newblock


\bibitem[\protect\citeauthoryear{Manola, Miller, McBride, et~al\mbox{.}}{Manola
  et~al\mbox{.}}{2004}]%
        {manola2004rdf}
\bibfield{author}{\bibinfo{person}{Frank Manola}, \bibinfo{person}{Eric
  Miller}, \bibinfo{person}{Brian McBride}, {et~al\mbox{.}}}
  \bibinfo{year}{2004}\natexlab{}.
\newblock \showarticletitle{RDF primer}.
\newblock \bibinfo{journal}{\emph{W3C recommendation}} \bibinfo{volume}{10},
  \bibinfo{number}{1-107} (\bibinfo{year}{2004}), \bibinfo{pages}{6}.
\newblock


\bibitem[\protect\citeauthoryear{McCarthy}{McCarthy}{1993}]%
        {mccarthy93contexts}
\bibfield{author}{\bibinfo{person}{John McCarthy}.}
  \bibinfo{year}{1993}\natexlab{}.
\newblock \showarticletitle{Notes on Formalizing Context}. In
  \bibinfo{booktitle}{\emph{Proceedings of the 13th International Joint
  Conference on Artificial Intelligence. Chamb{\'{e}}ry, France, August 28 -
  September 3, 1993}}. \bibinfo{publisher}{Morgan Kaufmann},
  \bibinfo{address}{Chamb{\'{e}}ry, France}, \bibinfo{pages}{555--562}.
\newblock
\urldef\tempurl%
\url{http://www-formal.stanford.edu/jmc/context3/context3.html}
\showURL{%
\tempurl}


\bibitem[\protect\citeauthoryear{Nagel, Affeldt, and Balke}{Nagel
  et~al\mbox{.}}{2021}]%
        {nagel2021datasetlinking}
\bibfield{author}{\bibinfo{person}{Denis Nagel}, \bibinfo{person}{Till
  Affeldt}, {and} \bibinfo{person}{Wolf{-}Tilo Balke}.}
  \bibinfo{year}{2021}\natexlab{}.
\newblock \showarticletitle{Data Narrations - Using flexible Data Bindings to
  support the Reproducibility of Claims in Digital Library Objects}. In
  \bibinfo{booktitle}{\emph{Proceedings of the Workshop on Digital
  Infrastructures for Scholarly Content Objects {(DISCO} 2021) co-located with
  {ACM/IEEE} Joint Conference on Digital Libraries 2021(JCDL 2021), Online (Due
  to the Global Pandemic), September 30, 2021}} \emph{(\bibinfo{series}{{CEUR}
  Workshop Proceedings}, Vol.~\bibinfo{volume}{2976})}.
  \bibinfo{publisher}{CEUR-WS.org}, \bibinfo{address}{Online},
  \bibinfo{pages}{19--23}.
\newblock
\urldef\tempurl%
\url{http://ceur-ws.org/Vol-2976/short-2.pdf}
\showURL{%
\tempurl}


\bibitem[\protect\citeauthoryear{Schardelmann and Otto}{Schardelmann and
  Otto}{2018}]%
        {SchardelmannOtto2018pollux}
\bibfield{author}{\bibinfo{person}{Tim Schardelmann} {and}
  \bibinfo{person}{Wolfgang Otto}.} \bibinfo{year}{2018}\natexlab{}.
\newblock \showarticletitle{POLLUX – von der Bedarfsanalyse zur technischen
  Umsetzung}.
\newblock \bibinfo{journal}{\emph{Bibliotheksdienst}} \bibinfo{volume}{52},
  \bibinfo{number}{3-4} (\bibinfo{year}{2018}), \bibinfo{pages}{225--234}.
\newblock
\urldef\tempurl%
\url{https://doi.org/10.1515/bd-2018-0029}
\showDOI{\tempurl}


\bibitem[\protect\citeauthoryear{Stoeckle, Witting, Kapadia, An, and
  Marks}{Stoeckle et~al\mbox{.}}{2022}]%
        {pmid34406670}
\bibfield{author}{\bibinfo{person}{K. Stoeckle}, \bibinfo{person}{B. Witting},
  \bibinfo{person}{S. Kapadia}, \bibinfo{person}{A. An}, {and}
  \bibinfo{person}{K. Marks}.} \bibinfo{year}{2022}\natexlab{}.
\newblock \showarticletitle{{{E}levated inflammatory markers are associated
  with poor outcomes in {C}{O}{V}{I}{D}-19 patients treated with remdesivir}}.
\newblock \bibinfo{journal}{\emph{J Med Virol}} \bibinfo{volume}{94},
  \bibinfo{number}{1} (\bibinfo{date}{01} \bibinfo{year}{2022}),
  \bibinfo{pages}{384--387}.
\newblock


\bibitem[\protect\citeauthoryear{Stoermer, Palmisano, Redavid, Iannone,
  Bouquet, and Semeraro}{Stoermer et~al\mbox{.}}{2006}]%
        {stoermer2006dlcontextualizedknowledgebase}
\bibfield{author}{\bibinfo{person}{Heiko Stoermer}, \bibinfo{person}{Ignazio
  Palmisano}, \bibinfo{person}{Domenico Redavid}, \bibinfo{person}{Luigi
  Iannone}, \bibinfo{person}{Paolo Bouquet}, {and} \bibinfo{person}{Giovanni
  Semeraro}.} \bibinfo{year}{2006}\natexlab{}.
\newblock \showarticletitle{Contextualization of a RDF Knowledge Base in the
  VIKEF Project}. In \bibinfo{booktitle}{\emph{Digital Libraries: Achievements,
  Challenges and Opportunities}}. \bibinfo{publisher}{Springer Berlin
  Heidelberg}, \bibinfo{address}{Berlin, Heidelberg},
  \bibinfo{pages}{101--110}.
\newblock
\showISBNx{978-3-540-49377-8}


\bibitem[\protect\citeauthoryear{Tchechmedjiev, Fafalios, Boland, Gasquet,
  Zloch, Zapilko, Dietze, and Todorov}{Tchechmedjiev et~al\mbox{.}}{2019}]%
        {tchechmedjiev2019claimskg}
\bibfield{author}{\bibinfo{person}{Andon Tchechmedjiev},
  \bibinfo{person}{Pavlos Fafalios}, \bibinfo{person}{Katarina Boland},
  \bibinfo{person}{Malo Gasquet}, \bibinfo{person}{Matth{\"a}us Zloch},
  \bibinfo{person}{Benjamin Zapilko}, \bibinfo{person}{Stefan Dietze}, {and}
  \bibinfo{person}{Konstantin Todorov}.} \bibinfo{year}{2019}\natexlab{}.
\newblock \showarticletitle{ClaimsKG: A Knowledge Graph of Fact-Checked
  Claims}. In \bibinfo{booktitle}{\emph{The Semantic Web -- ISWC 2019}}.
  \bibinfo{publisher}{Springer International Publishing},
  \bibinfo{address}{Cham}, \bibinfo{pages}{309--324}.
\newblock
\showISBNx{978-3-030-30796-7}


\bibitem[\protect\citeauthoryear{Thompson, Karunadasa, Varma, Schoenwaelder,
  and Clements}{Thompson et~al\mbox{.}}{2021}]%
        {pmid34729931}
\bibfield{author}{\bibinfo{person}{C. Thompson}, \bibinfo{person}{H.
  Karunadasa}, \bibinfo{person}{D. Varma}, \bibinfo{person}{M. Schoenwaelder},
  {and} \bibinfo{person}{W. Clements}.} \bibinfo{year}{2021}\natexlab{}.
\newblock \showarticletitle{{{I}mpact of {C}{O}{V}{I}{D} vaccination rollout on
  the use of computed tomography venography for the assessment of cerebral
  venous sinus thrombosis}}.
\newblock \bibinfo{journal}{\emph{J Med Imaging Radiat Oncol}}
  \bibinfo{volume}{65}, \bibinfo{number}{7} (\bibinfo{date}{Dec}
  \bibinfo{year}{2021}), \bibinfo{pages}{883--887}.
\newblock


\bibitem[\protect\citeauthoryear{Vrande{\v{c}}i{\'{c}} and
  Kr{\"{o}}tzsch}{Vrande{\v{c}}i{\'{c}} and Kr{\"{o}}tzsch}{2014}]%
        {vrandevcic2014wikidata}
\bibfield{author}{\bibinfo{person}{Denny Vrande{\v{c}}i{\'{c}}} {and}
  \bibinfo{person}{Markus Kr{\"{o}}tzsch}.} \bibinfo{year}{2014}\natexlab{}.
\newblock \showarticletitle{{Wikidata: a free collaborative knowledgebase}}.
\newblock \bibinfo{journal}{\emph{Commun. ACM}} \bibinfo{volume}{57},
  \bibinfo{number}{10} (\bibinfo{year}{2014}), \bibinfo{pages}{78--85}.
\newblock


\bibitem[\protect\citeauthoryear{Wilkinson, Dumontier, Aalbersberg, Appleton,
  Axton, Baak, and et~al.}{Wilkinson et~al\mbox{.}}{2016}]%
        {wilkinson2016fairprincles}
\bibfield{author}{\bibinfo{person}{Mark~D. Wilkinson}, \bibinfo{person}{Michel
  Dumontier}, \bibinfo{person}{IJsbrand~Jan Aalbersberg},
  \bibinfo{person}{Gabrielle Appleton}, \bibinfo{person}{Myles Axton},
  \bibinfo{person}{Arie Baak}, {and} \bibinfo{person}{et al.}}
  \bibinfo{year}{2016}\natexlab{}.
\newblock \showarticletitle{The FAIR Guiding Principles for scientific data
  management and stewardship}.
\newblock \bibinfo{journal}{\emph{Scientific Data}} \bibinfo{volume}{3},
  \bibinfo{number}{1} (\bibinfo{date}{15 Mar} \bibinfo{year}{2016}),
  \bibinfo{pages}{160018}.
\newblock
\showISSN{2052-4463}
\urldef\tempurl%
\url{https://doi.org/10.1038/sdata.2016.18}
\showDOI{\tempurl}


\bibitem[\protect\citeauthoryear{{Wylot}, {Cudré-Mauroux}, {Hauswirth}, and
  {Groth}}{{Wylot} et~al\mbox{.}}{2017}]%
        {ProvenanceInLinkedData}
\bibfield{author}{\bibinfo{person}{M. {Wylot}}, \bibinfo{person}{P.
  {Cudré-Mauroux}}, \bibinfo{person}{M. {Hauswirth}}, {and}
  \bibinfo{person}{P. {Groth}}.} \bibinfo{year}{2017}\natexlab{}.
\newblock \showarticletitle{Storing, Tracking, and Querying Provenance in
  Linked Data}.
\newblock \bibinfo{journal}{\emph{IEEE Transactions on Knowledge and Data
  Engineering}} \bibinfo{volume}{29}, \bibinfo{number}{8}
  (\bibinfo{year}{2017}), \bibinfo{pages}{1751--1764}.
\newblock


\end{thebibliography}

\end{document}